\documentclass[lettersize,journal]{IEEEtran}
\usepackage{amsmath,amsfonts}
\usepackage{algorithm}
\usepackage{array}
\usepackage{multirow}
\usepackage{subfigure}
\usepackage{textcomp}
\usepackage{stfloats}
\usepackage{url}
\usepackage{verbatim}
\usepackage{graphicx}
\usepackage{cite}
\usepackage{caption}
\usepackage{amssymb}
\usepackage{algpseudocode}
\usepackage{soul}    
\usepackage{xcolor}  
\usepackage{array}
\usepackage{booktabs} 
\usepackage{colortbl} 
\usepackage{array} 
\definecolor{lightgray}{gray}{0.9} 
\usepackage{booktabs}
\usepackage{siunitx}
\usepackage{multirow}


\begin{document}

\title{Conquering High Packet-Loss Erasure: MoE Swin Transformer-Based Video Semantic Communication}

\author{\relax Lei Teng, Senran Fan, Chen Dong\IEEEauthorrefmark{1}, Haotai Liang, Zhicheng Bao,\\ Xiaodong Xu, \textit{Senior Member, IEEE}, Rui Meng, Ping Zhang, \textit{Fellow, IEEE}}

\markboth{Journal of \LaTeX\ Class Files,~Vol.~14, No.~8, August~2021}%
{Shell \MakeLowercase{\textit{et al.}}: A Sample Article Using IEEEtran.cls for IEEE Journals}


\maketitle

\begin{abstract}

Semantic communication with joint semantic-channel coding robustly transmits diverse data modalities but faces challenges in mitigating semantic information loss due to packet drops in packet-based systems. Under current protocols, packets with errors are discarded, preventing the receiver from utilizing erroneous semantic data for robust decoding. To address this issue, a packet-loss-resistant MoE Swin Transformer-based Video Semantic Communication (MSTVSC) system is proposed in this paper. Semantic vectors are encoded by MSTVSC and transmitted through upper-layer protocol packetization. To investigate the impact of the packetization, a theoretical analysis of the packetization strategy is provided. To mitigate the semantic loss caused by packet loss, a 3D CNN at the receiver recovers missing information using un-lost semantic data and an packet-loss mask matrix. Semantic-level interleaving is employed to reduce concentrated semantic loss from packet drops. To improve compression, a common-individual decomposition approach is adopted, with downsampling applied to individual information to minimize redundancy. The model is lightweighted for practical deployment. Extensive simulations and comparisons demonstrate strong performance, achieving an MS-SSIM greater than 0.6 and a PSNR exceeding 20 dB at a 90\% packet loss rate.


\end{abstract}


\section{Introduction}
\IEEEPARstart{W}{ith} the rise of the Internet, vast amounts of data and complex communication environments have imposed higher demands on the transmission efficiency and robustness of communication systems \cite{ref1}. However, the transmission capacity of traditional communication systems at the physical layer is increasingly approaching the Shannon limit. Meanwhile, the lossless communication architecture based on entropy coding and channel coding faces the cliff effect in harsh communication environments, highlighting the urgent need for an emerging communication technology that offers high robustness within limited bandwidth.

Weaver \cite{ref2}, one year after Shannon proposed information theory \cite{ref3}, identified three levels of communication: syntactic, semantic, and pragmatic. Traditional communication has primarily been focused on the syntactic level, whereas, with the rapid development of artificial intelligence, semantic communication based on neural networks, exhibiting strong noise resistance, has gradually emerged and attracted widespread attention \cite{ref4}-\cite{ref6}. Unlike traditional communication, semantic communication is primarily concerned with the semantics underlying the transmitted information. In other words, semantic communication can tolerate certain syntactic-level errors, such as bit errors, provided that semantic distortion remains minimal. Consequently, semantic communication can achieve greater robustness while increasing compression rates, thereby reducing demands on the communication environment, alleviating transmission pressure, and enhancing transmission efficiency.

Joint source-channel coding (JSCC) based on deep learning, which accounts for channel noise in both source coding and channel coding, exhibits strong robustness against the cliff effect, distinguishing it among numerous AI technologies and leading to its widespread application in semantic communication research. In the domain of image sources, a layer-based semantic communication system for images (LSCI) is introduced \cite{ref8}. It posits that semantic transmission is essentially the dissemination of artificial intelligence models and employs semantic slice-models (SeSM) to achieve this. Inspired by model division multiple access (MDMA) \cite{ref9}, a novel method for efficiently and controllably transmitting video data over noisy wireless channels is proposed \cite{ref10}, utilized for extracting common semantic features from video frames for video compression. Furthermore, with the advancement of semantic communication, additional studies have emerged across various source types, such as text \cite{ref7}, \cite{ref11}, speech \cite{ref12}\cite{ref13}, images \cite{ref14} -\cite{ref17}, video \cite{Qi2024} -\cite{ref21}, VR \cite{ref22}, and 3D point clouds \cite{ref23}. Previous work assumed that any complex value could be transmitted through the channel. However, if directly applied to digital communication, a full-resolution digital modulation and demodulation system would be required, which is evidently challenging to implement. To address this issue, a semantic digital modulation constellation mapping (sDMCM) scheme based on pulse amplitude modulation (PAM) / quadrature amplitude modulation (QAM), considering the internal correlation of semantic information, is proposed in \cite{ref26}.

Although these studies collectively address the application of semantics in digital communication systems at the physical layer, the impact of existing upper-layer protocols on semantic digital communication has been overlooked. Currently, widely adopted communication protocols, such as TCP and UDP, are packet-based, where data is divided into multiple packets before being transmitted through the channel. However, most existing semantic communication system studies focus on direct modulation mapping at the physical layer, neglecting the packet handling in upper-layer protocols. Until new communication standards emerge, integrating error-tolerant reception systems compatible with semantic communication into commercial frameworks remains challenging, significantly undermining the noise-resistant capabilities of JSCC. Specifically, packets containing errors are discarded entirely, even though the decoding end, trained with JSCC, could potentially handle such errors. Consequently, there is an urgent need for a native AI semantic communication system that can be applied within existing protocol frameworks and is capable of mitigating semantic erasure channels, namely packet loss channels.

Previous semantic communication research predominantly relied on CNNs as the backbone, but a shift toward a new backbone is emerging. In \cite{ref27}, the Swin Transformer is introduced as a versatile backbone for computer vision, leveraging a hierarchical window-based self-attention mechanism to efficiently integrate local and global feature modeling with low computational complexity and superior performance across vision tasks. An image semantic communication system based on the Swin Transformer is proposed in \cite{ref25}, outperforming CNN-based systems in reconstruction performance. To handle video sources, the 3D Swin Transformer is developed in \cite{ref28}, enhancing video understanding by incorporating a temporal dimension into the window-based self-attention mechanism, thus effectively capturing spatiotemporal features. Amid the rise of large-scale models, the Mixture of Experts (MoE) model \cite{ref29} has gained traction for improving performance, reducing computational costs, and enabling efficient scaling. In video semantic encoding and decoding, MoE optimizes computational resources through dynamic expert allocation, enhancing the modeling of complex spatiotemporal features and significantly boosting video processing performance. A comparison of the codec in this paper with those in existing studies is provided in Tab. \ref{Review}, highlighting similarities and differences.

\begin{table}[h]
	\centering
	\caption{Semantic Video Communication Review Table}\label{Review}
	\begin{tabular}{>{\centering\arraybackslash}p{2cm}|c|>{\centering\arraybackslash}p{4cm}}
		\toprule
		\textbf{Backbone} & \textbf{Reference} & \textbf{Codec Features} \\
		\midrule
		
		\multirow{5}{*}{CNN} & \cite{ref10} & Common and personalized feature separation \\ \cline{2-3}
		& \cite{Qi2024}, \cite{Yang2025} & Frame interpolation \\ 
		\cline{2-3}
		& \cite{Jiang2023}, \cite{ref19} & Only keypoint transmission \\ 
		\cline{2-3}
		& \cite{Samarathunga2024} \cite{ref21} & Keyframe and residual frame \\ 
		\cline{2-3}
		& \cite{Tong2025} & Multimodal redundant compression \\ 
		\cline{1-3}
		\multirow{2}{*}{Swin Transformer} & \cite{Dai2023} & Entropy compression \\ 
		\cline{2-3}
		& \cite{ref20} & Context-driven semantic feature modeling \\ \cline{2-3}
		& \cite{Tian2025} & Multi-modal \\ 
		\cline{1-3}
		3D Swin Transformer + MoE & Our paper & Personalized feature downsampling after common and personalized feature separation \\ 
		\bottomrule
	\end{tabular}
\end{table}

To realize a next-generation native AI communication system and bridge the gap between existing semantic digital communication systems and current digital communication protocols, this semantic video communication system is trained and designed specifically for packet-loss channels, unlike previous AWGN and fading channels. An additional packet-loss recovery module is designed at the receiver to generate missing information. The main contributions of this paper are summarized as follows:

\begin{enumerate}
	\item A packet-loss-resistant semantic video communication system based on the Swin Transformer is proposed. Videos are divided into Groups of Pictures (GOPs), encoded into semantic vectors by a semantic encoder, and transmitted via packetization. At the receiver, lost information is predicted and reconstructed based on un-lost data, ensuring available video communication under packet loss conditions, with discernible frames preserved even at high packet loss rates.
	\item  A packetization strategy based on semantics is theoretically investigated in this paper, analyzing the impact of packet length on packet loss rate under different symbol error rates (SER). Based on the relationship curve between packet loss rate and semantic performance, the impact of packet length on semantic performance is derived. On this basis, packet length can be flexibly adjusted according to receiver performance requirements under a fixed SER, followed by reducing the number of packets to reduce head information redundancy. This enables a trade-off between redundancy and robustness under varying SER conditions, approaching the performance required by the receiver.
	\item A packet-loss-resistant method is proposed to counter semantic loss caused by packet drops, effectively integrating the semantic communication paradigm with existing packet-based digital communication systems. Interleaving is employed to mitigate concentrated semantic information loss due to packet drops. Additionally, a 3D CNN network is designed at the receiver to generate and recover missing information using un-lost semantic data and a packet-loss mask matrix.
	\item High-quality video communication with high compression is achieved through neural network model design. Semantic vectors are decomposed into common and individual feature, with individual feature further downsampled and compressed to reduce redundancy. A temporal semantic information codec based on the Mixture of Experts (MoE) 3D Swin Transformer is designed, allowing automatic selection of the optimal expert for encoding and decoding different image patch.
	\item  A semantic codec based on spatial-temporal compression separation is designed, where spatial source compression is first applied to each frame in the input GOP, followed by temporal and spatial information compression of the semantic information from consecutive compressed video frames, enhancing compression efficiency.
	\item Extensive simulations and performance comparisons are conducted.
\end{enumerate}

The subsequent sections of the paper are planned as follows:
In the  section II, the system model is primarily introduced, including an overview of the overall architecture of the packet-loss digital communication system, semantic-level interleaving and segmentation.
The section III derives the relationship between SER, packet length, and packet loss rate.
The section IV focuses on presenting the network framework of the proposed MSTVSC.
The section V presents the simulation results and performance comparisons.
The section VI provides a summary of the entire paper.

\section{System Model}
\begin{figure*}[!htbp]
	\centering
	\includegraphics[width=0.99\textwidth]{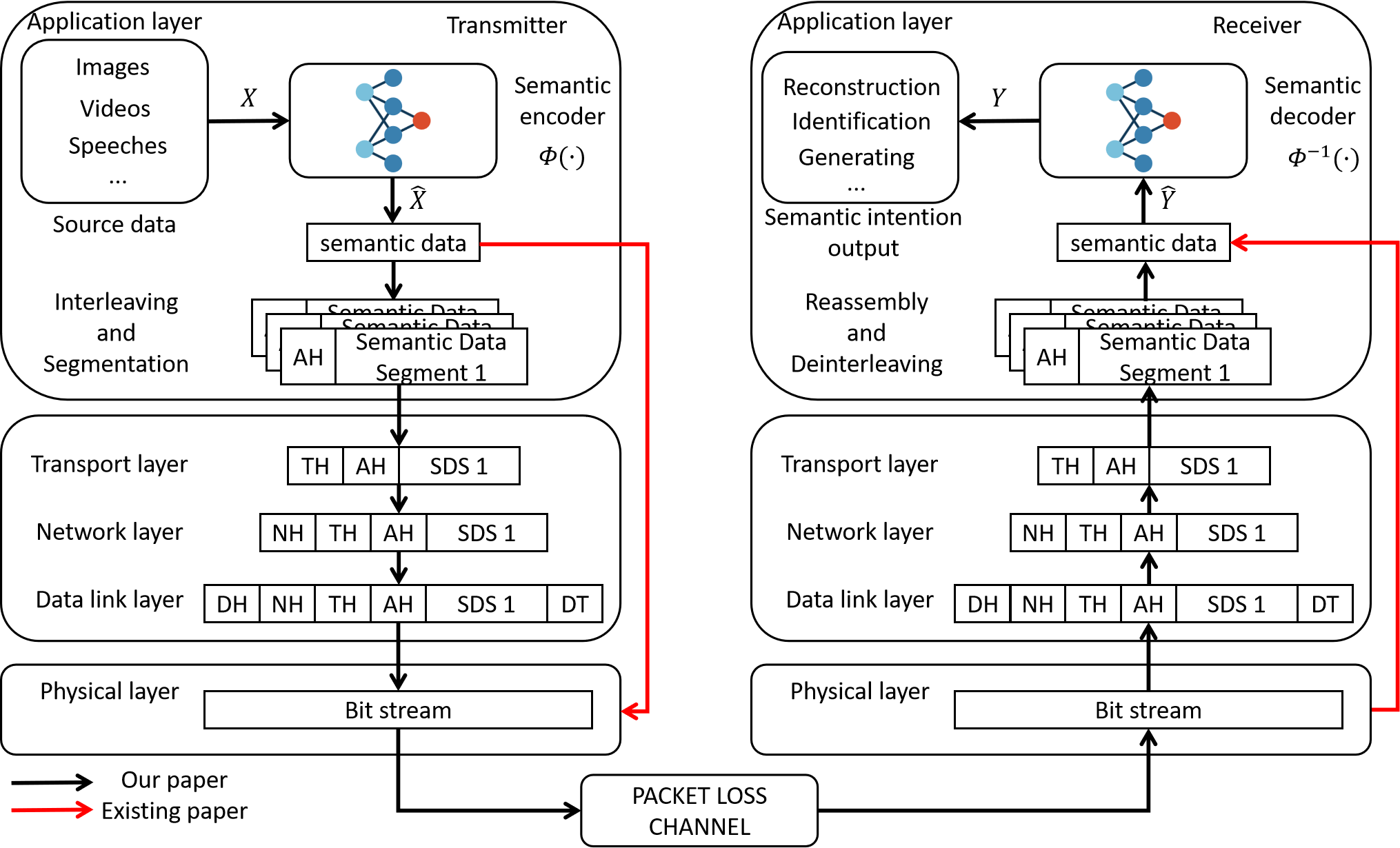}\\

	\caption{\footnotesize The semantic digital communication system model proposed in this paper is depicted by the black arrow in the figure. At the transmitter, source data is encoded by the semantic encoder to generate semantic data. This data is interleaved to form a disordered semantic data vector, which is then segmented into semantic data segments (SDS). An application-layer header (AH) is appended to each segment. The data subsequently passes through the transport, network, and data link layers, where a transport-layer header (TH), network-layer header (NH), data link-layer header (DH), and data link-layer trailer (DT) are added, respectively. The data is then transmitted through a packet-loss channel via the physical layer. The receiver performs the inverse process. In contrast, as shown by the red arrow, existing physical-layer semantic communication studies bypass the transport, network, and data link layers, directly transmitting semantic data through the physical layer.}
	\label{system model}
\end{figure*}

A packet-based semantic digital communication system, is proposed and depicted by the black arrow in Fig. \ref{system model}. The system is structured into application, transport, network, data link, and physical layers. At the transmitter’s application layer, source data $X$ is processed by the semantic encoder $\Phi(\cdot)$ to yield semantic data $\hat{X}$, which is then interleaved and segmented before being passed to lower layers. At the receiver’s application layer, data from lower layers is reassembled and de-interleaved to produce $\hat{Y}$, which is decoded by the semantic decoder $\Phi^{-1}(\cdot)$ to generate the semantic intent output. Details of interleaving, de-interleaving, segmentation, and reassembly are provided in subsections A and B. The transport, network, and data link layers adhere fully to commercial protocols, ensuring compatibility with most existing communication devices. For the packet-loss-resistant semantic video communication system examined, further design is based on UDP. Unlike TCP, which retransmits lost packets to ensure data integrity, UDP’s connectionless, lightweight nature prioritizes low-latency transmission without reliability mechanisms, aligning with semantic communication’s tolerance for bit-level errors. This makes UDP preferable for real-time video communication, where low latency is critical, and its smaller header (8 bytes versus TCP’s 20 bytes) improves bandwidth efficiency. TCP’s retransmission and connection establishment, while suitable for high-integrity applications like file transfers, increase latency, making it less ideal for real-time scenarios. In the physical layer, existing channel coding and decoding are disabled to reduce redundancy, as the semantic codec, trained for packet-loss channels, tolerates some packet loss and does not require a lossless bitstream.

\subsection{Interleaving}
\begin{figure}[!htbp]
	\centering
	\includegraphics[width=0.49\textwidth]{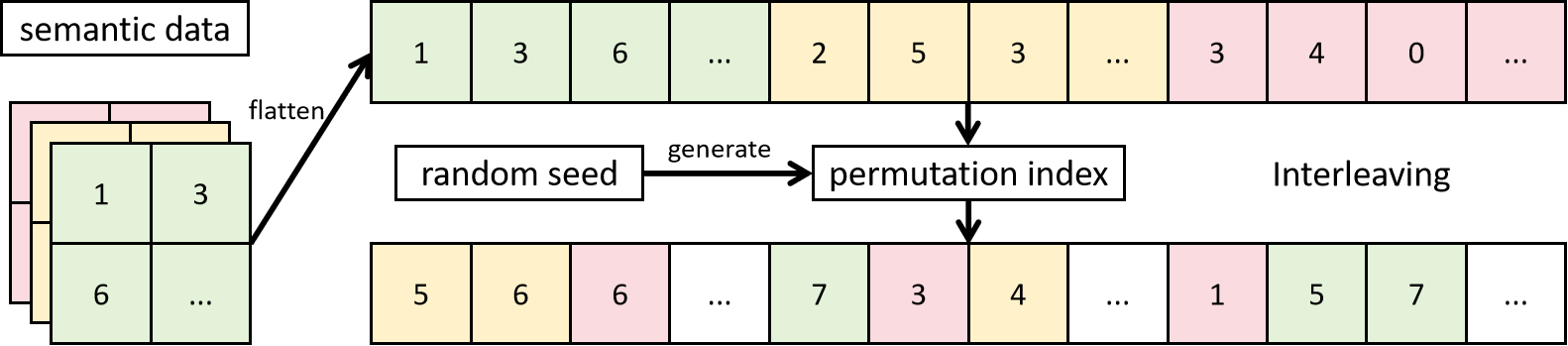}\\
	\caption{\footnotesize A schematic diagram of application-layer semantic information element interleaving. In the semantic data on the left, different colors represent the outputs of different channels in the semantic information matrix. Initially, the semantic data is flattened into a one-dimensional vector. Subsequently, a permutation index of the same length as the semantic data is generated based on a specified random seed. Interleaving is then performed according to the permutation index, yielding the interleaved result.}
	\label{Interleaving}
\end{figure}

 Unlike bit-level physical-layer interleaving, semantic communication prioritizes minimizing semantic distortion. As highlighted in \cite{ref26}, semantic distortion is closely linked to changes in the semantic information matrix elements, not bits. Thus, an interleaving scheme is developed to disperse highly correlated adjacent semantic information into a low-correlation matrix, preventing consecutive semantic information loss due to packet drops during transmission. This enhances video reconstruction performance at the receiver. A random index generation method is employed to maximize the dispersion of correlated semantic information. To eliminate the overhead of transmitting indices and the need for an error-free signaling channel, a pre-agreed random seed is used by the transmitter and receiver, ensuring consistent interleaving indices. This approach also provides information security, as intercepted data cannot be decoded without the interleaving indices, even with the same decoder. The interleaving process is illustrated in Fig. \ref{Interleaving}, while the de-interleaving process, being its inverse, is omitted to avoid redundancy.

\subsection{Segmentation}
\begin{figure}[!htbp]
	\centering
	\includegraphics[width=0.49\textwidth]{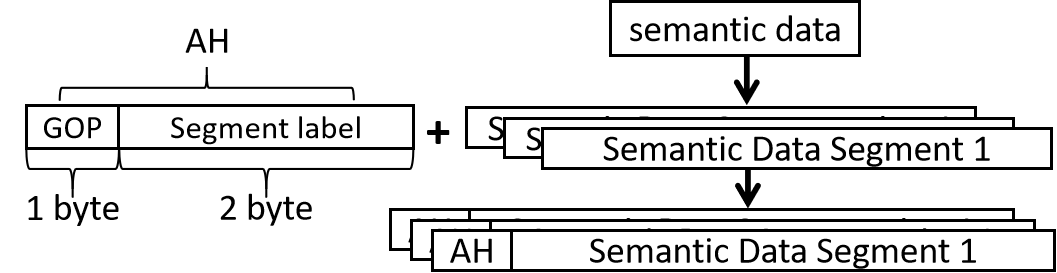}\\
	\caption{\footnotesize Application layer segmentation flowchart.}
	\label{Segmentation}
\end{figure}

In UDP, data segmentation is not conducted at the transport layer, so if semantic data is directly processed by UDP, any symbol error results in the entire packet being discarded, causing complete loss of semantic data. However, the part of error-free semantic information can still be used for decoding video semantic data. To address this, semantic data is segmented at the application layer before transmission via UDP. This ensures that errors in individual segments lead only to those segments being discarded, preserving the remaining data. Each segment is assigned an application-layer header (AH) to facilitate reassembly at the receiver. As depicted in the segmentation flowchart (Fig. \ref{Segmentation}), the AH structure comprises: the first byte, indicating the current GOP number, and the second and third bytes, representing the segment label that denotes the segment’s position within the GOP’s semantic data. The GOP number enables the receiver to identify whether received data belongs to the same GOP. Assuming no out-of-order reception, a change in the received GOP number compared to the buffer’s GOP number indicates that the previous GOP’s semantic data has been fully received and can be forwarded to the semantic decoder for video decoding. Using the segment labels, segments are reassembled into complete semantic data, with missing segments replaced by zeros, even if some are lost during transmission.


\section{Theoretical Derivation}

Existing semantic communication research primarily examines how physical channel parameters, such as SNR in AWGN channels, affect semantic performance at the receiver. However, in packet-based digital communication, packets with erroneous bits are discarded, leading to packet loss. Until error-tolerant transmission protocols tailored for semantic communication are developed, relying solely on physical layer designs limits the realization of semantic communication’s full potential. Consequently, the impact of packet loss rate on semantic performance in digital communication is investigated in this study. The relationship between packet loss rate \(P_L\) and parameters, including data payload length \(L_{data}\), combined header and trailer length \(L_{head}\), packet length \(P_{length}\), number of packets \(N_p\), modulation order \(2^M\), symbol error rate (SER) \(P_s\), and original data length \(L_{origin}\), is derived, based on the packet encapsulation process illustrated in Fig. \ref{system model}. By fitting the experimental relationship curve \(f_p\) between packet loss rate and semantic performance, the connection between these parameters and semantic performance is established, guiding system optimization. Substituting \(P_L\) into \(f_p\) enables the relationship between semantic performance and communication parameters to be quantified, allowing the minimization of total data transmission \(L_{total}\) while meeting the receiver’s semantic performance threshold \(\Gamma_{P}\).

The packet loss rate can be expressed as follows:

\begin{equation}
	\begin{split}
		P_L = 1 - (1 - P_s)^{N_{symbol}},
	\end{split}
\end{equation}
where \(N_{symbol}\) represents the total number of transmitted symbols in a packet, which can be calculated as:

\begin{equation}
	\begin{split}
		N_{symbol} = \frac{8 \cdot P_{length}}{M},
	\end{split}
\end{equation}
where, since \(P_{length}\) is in bytes, it is multiplied by 8 to convert the unit to bits, and then divided by the modulation order parameter \(M\) to determine the number of symbols required to transmit a packet. Furthermore,
\begin{equation}
	\begin{split}
		P_{length} = L_{head} + L_{data}.
	\end{split}
\end{equation}
Given the original data length \(L_{origin}\),
\begin{equation}
	\begin{split}
		L_{data} = \frac{L_{origin}}{N_p}.
	\end{split}
\end{equation}
In this case, the total transmitted data amount \(L_{total}\) is:
\begin{equation}
	\begin{split}
		L_{total}& = N_p \cdot P_{length}\\
		& = N_p \cdot (L_{head} + L_{data})\\
		& = N_p \cdot L_{head} + L_{origin},
		\label{Lt}
	\end{split}
\end{equation}

Combining the above, the packet loss rate can ultimately be expressed as:
\begin{equation}
	\begin{split}
		P_L &= 1 - (1 - P_s)^{\frac{8 \cdot (L_{head} + \frac{L_{origin}}{N_p})}{M}}\\
		&=1 - (1 - P_s)^{\frac{8 \cdot (\frac{L_{total}}{N_p})}{M}}.
	\end{split}
\end{equation}

Based on experimental simulations, it is determined that \(f_p\) is a monotonically decreasing function of \(P_L\), with its maximum value \(\Gamma_{MAX}\) at \(P_L = 0\). For a given \(P_s\), the minimum packet loss rate is:
\begin{equation}
	\begin{split}
		P_{L_{min}}(P_s) > 1 - (1 - P_s)^{\frac{8 \cdot L_{head}}{M}},
	\end{split}
\end{equation}
where \(P_{L_{min}}(P_s)\) represents the minimum \(P_L\) under a given \(P_s\). Consequently, the maximum performance \(\Gamma_{MAX}(P_s)\) can be expressed as:
\begin{equation}
	\begin{split}
		\Gamma_{MAX}(P_s) = f_p\left(1 - (1 - P_s)^{\frac{8 \cdot L_{head}}{M}}\right).
	\end{split}
\end{equation}

Furthermore, when \(L_{head}\), \(L_{origin}\), \(0\leq P_s\leq1\), and \(M\) are fixed, the derivative of Equation (23) with respect to \(N_p\) is:

\begin{equation}
	\begin{split}
		\frac{dP_L}{dN_p} = \frac{8 L_{origin} \ln(1 - P_s)}{M N_p^2} \cdot (1 - P_s)^{\frac{8 \cdot (L_{head} + \frac{L_{origin}}{N_p})}{M}} < 0,
	\end{split}
\end{equation}
indicating that \(P_L\) decreases monotonically with \(N_p\). Since Eq. (\ref{Lt}) indicates that \(L_{total}\) increases monotonically with \(N_p\), the following mathematical relationship holds:
\begin{equation}
	\begin{split}
		\forall \,\, \Gamma_{P} \leq \Gamma_{\text{MAX}}(P_s),\,\, \exists \,\, N_{p_{min}}\,\, \text{s.t}.\,\, f_p(P_L) \geq \Gamma_{P}
	\end{split}
\end{equation}

\section{The Proposed MSTVSC Framework}
\begin{figure*}[!htbp]
	\centering
	\includegraphics[width=0.99\textwidth]{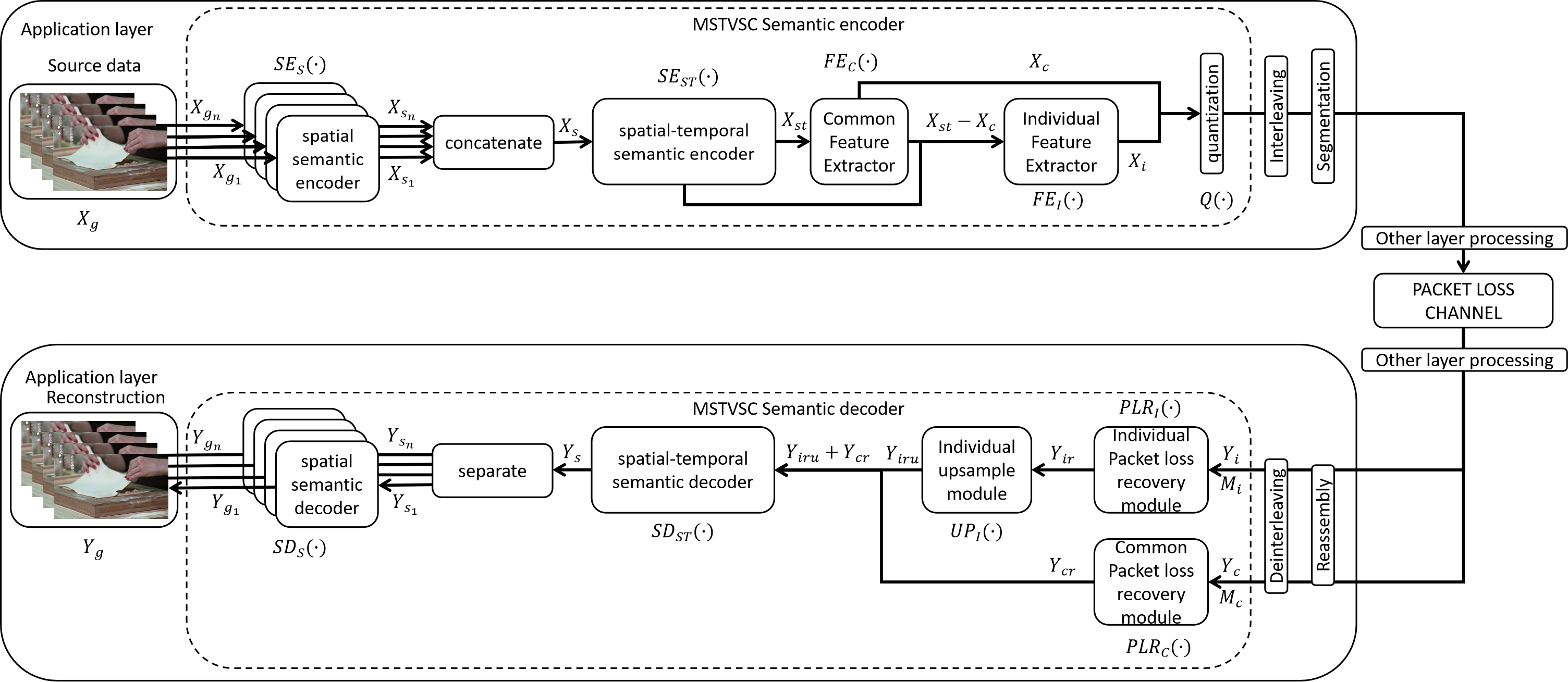}\\
	\caption{\footnotesize The overall architecture of our MSTVSC for wireless image transmission.}
	\label{MSTVSC}
\end{figure*}

The MSTVSC architecture for packet-loss-resistant transmission is outlined in Fig. \ref{MSTVSC}. At the transmitter, video source data is processed by the MSTVSC encoder, generating quantized semantic data. This data undergoes interleaving and segmentation at the application layer before being forwarded to lower layers for processing and transmission over a packet-loss channel. At the receiver, semantic data with missing information due to packet loss is processed by lower layers, then reassembled and de-interleaved at the application layer. The data is subsequently decoded by the MSTVSC decoder to reconstruct the video. The detailed encoding and decoding processes of MSTVSC are described below:

\subsection{The spatial semantic codec}
\begin{figure}[!htbp]
	\centering
	\includegraphics[width=0.5\textwidth]{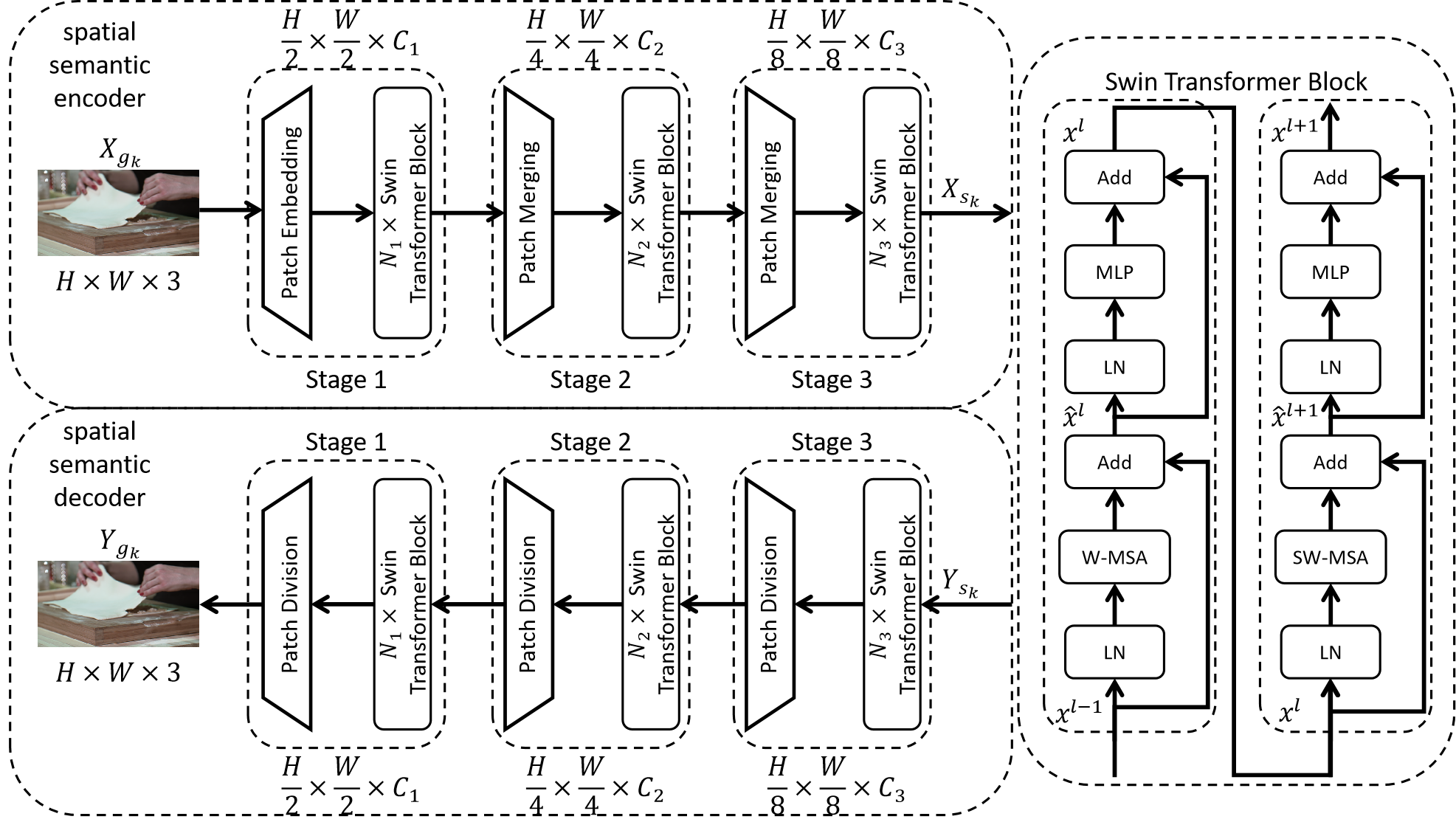}\\
	\caption{\footnotesize The architecture of the spatial semantic code.}
	\label{spatial semantic encoder}
\end{figure}

The structure of the spatial semantic codec is presented in Fig. \ref{spatial semantic encoder}, with its core architecture built upon the Swin Transformer as described in \cite{ref27}. Within the spatial semantic encoder, a single frame \(X_{g_k} \in \mathbb{R}^{H \times W \times 3}\) from the GOP is divided into \(H/2 \times W/2\) non-overlapping patches. These patches are handled by the Patch Embedding module to produce initial embedding tokens. These tokens are then processed through \(N_1\) Swin Transformer blocks to extract deeper semantic features. The combination of the Patch Embedding module and the \(N_1\) Swin Transformer blocks forms what is termed ``Stage 1.'' As depicted on the right side of Fig. \ref{spatial semantic encoder}, the Swin Transformer block employs a Multi-Head Self-Attention (MSA) module alongside a feedforward network, as outlined in \cite{ref27}, to derive semantic information from the patches. The shifted-window-based attention mechanism supports the modeling of long-range dependencies by partitioning the image into a grid of windows, where self-attention is applied locally within each window.

To achieve higher compression efficiency and obtain more concise semantic representations, patch merging layers and deeper architectures are employed. In Stage 2, the adjacent embeddings from Stage 1 are fused through a patch merging operation, reducing the concatenated embeddings from a dimension of \(4C_1\) to \(C_2\). The resulting tokens, with a resolution of \(H/4 \times W/4\), are subsequently processed by \(N_2\) Swin Transformer blocks. Stage 3 adopts a comparable approach, incorporating downsampling patch merging layers followed by \(N_3\) Swin Transformer blocks. This hierarchical deepening of the network enhances its ability to model long-range dependencies, incorporate global context, and capture intricate details in high-resolution images, ultimately yielding the semantic information vector \(X_{s_k}\) for a single frame. In the spatial semantic decoder, the input consists of the semantic information vector for a single frame, defined as \(Y_{s_k} = Y_{iru} + Y_{cr}\), with each stage designed as the inverse of its corresponding encoder stage, details of which are omitted here.

\subsection{The spatial-temporal semantic codec}
\begin{figure*}[!htbp]
	\centering
	\includegraphics[width=0.79\textwidth]{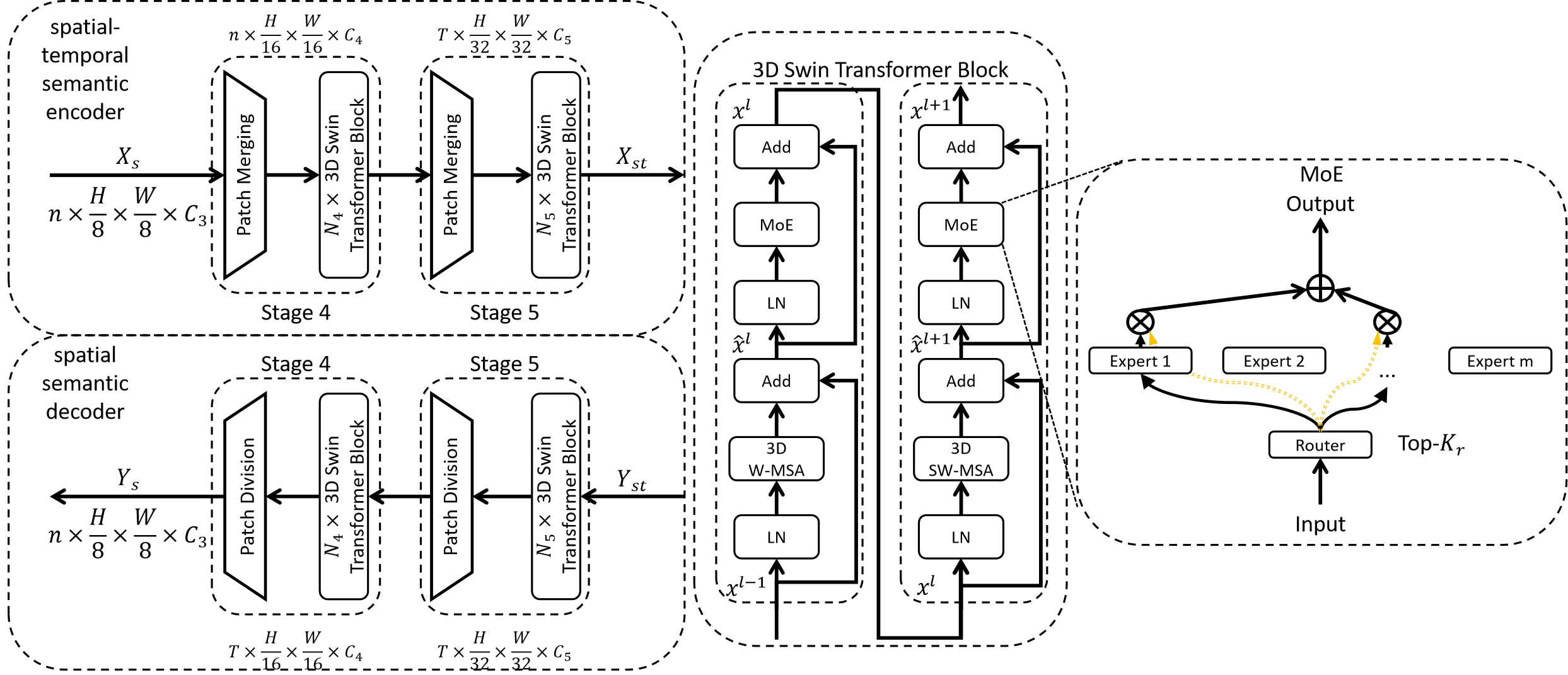}\\
	\caption{\footnotesize The architecture of the spatial-temporal semantic codec.}
	\label{spatial time semantic encoder}
\end{figure*}
To capture the correlations across different time steps within a GOP, a spatial-temporal semantic codec is designed, as illustrated in Fig. \ref{spatial time semantic encoder}. The stage design is similar to that of the spatial semantic codec, with each stage comprising a patch merging layer and several Swin Transformer blocks. However, to incorporate temporal dimension information, the standard Swin Transformer is replaced with a 3D Swin Transformer \cite{ref28}, which extends the shifted 2D window mechanism of the Swin Transformer to 3D windows. This introduces cross-window connections while maintaining the efficient computation of self-attention based on non-overlapping windows. To fully account for the temporal dimension information of the entire GOP, the temporal dimension of the window size is consistently set to \(n\), where \(n\) represents the number of frames in the GOP. Additionally, to better capture diverse spatial-temporal semantic information, the simple MLP layer in the original 3D Swin Transformer design is modified into a Mixture of Experts (MoE) layer \cite{ref29}. The specific MoE architecture is depicted on the right side of Fig. \ref{spatial time semantic encoder}, where the input is processed by a router to obtain weight vectors for \(m\) experts. The top \(K_r\) experts are selected to process the input, and their outputs are weighted and summed to produce the final output.

\subsection{The Common Feature Extractor and the Individual Feature Extractor}
\begin{figure}[!htbp]
	\centering
	\includegraphics[width=0.49\textwidth]{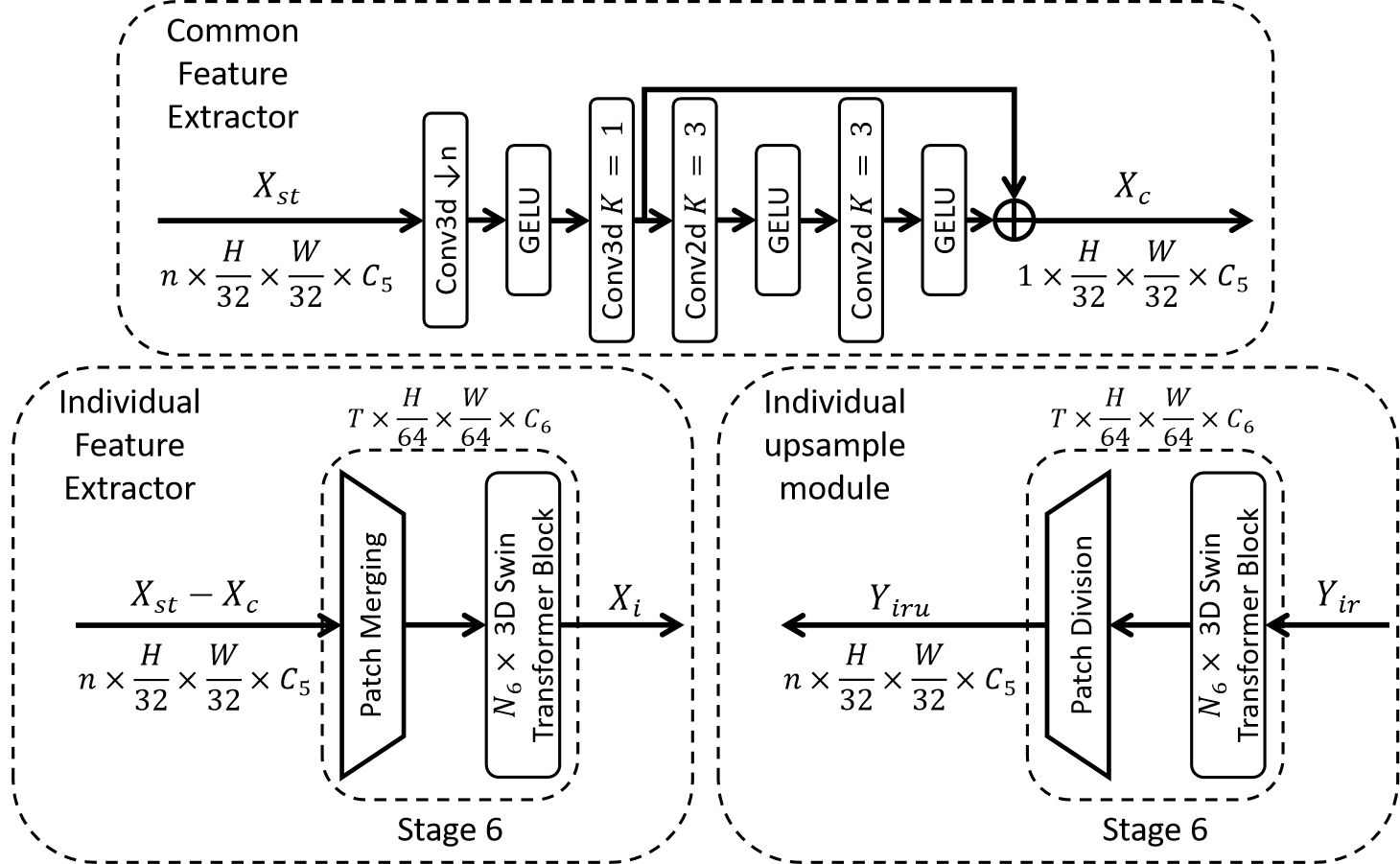}\\
	\caption{\footnotesize The architecture of the Common Feature Extractor and the Individual Feature Extractor.}
	\label{Feature Extractor}
\end{figure}
To further compress the redundancy of common information in the temporal dimension, the Common Feature Extractor and Individual Feature Extractor are designed as shown in Fig. \ref{Feature Extractor}. In the Common Feature Extractor, the input spatial-temporal semantic information vector \(X_{st}\) is first processed by a 3D convolutional layer, which downsamples the temporal dimension by a factor of \(n\), effectively compressing the temporal dimension to 1. This is followed by an activation function and a 3D convolutional layer with a \(1 \times 1 \times 1\) kernel to enhance information exchange between channels. The final common feature vector \(X_c\) is then obtained through a convolutional residual network. Subsequently, the preliminary individual features are derived by subtracting \(X_c\) from \(X_{st}\), and these are input into the Individual Feature Extractor for further feature compression. The stage framework of the Individual Feature Extractor is identical to that of the spatial-temporal semantic encoder and will not be elaborated further. Additionally, Fig. \ref{Feature Extractor} illustrates the Individual Upsample Module at the decoding end, which performs the inverse process of the Individual Feature Extractor to decode the compressed individual features.

\subsection{The Packet loss recovery module}
\begin{figure}[!htbp]
	\centering
	\includegraphics[width=0.4\textwidth]{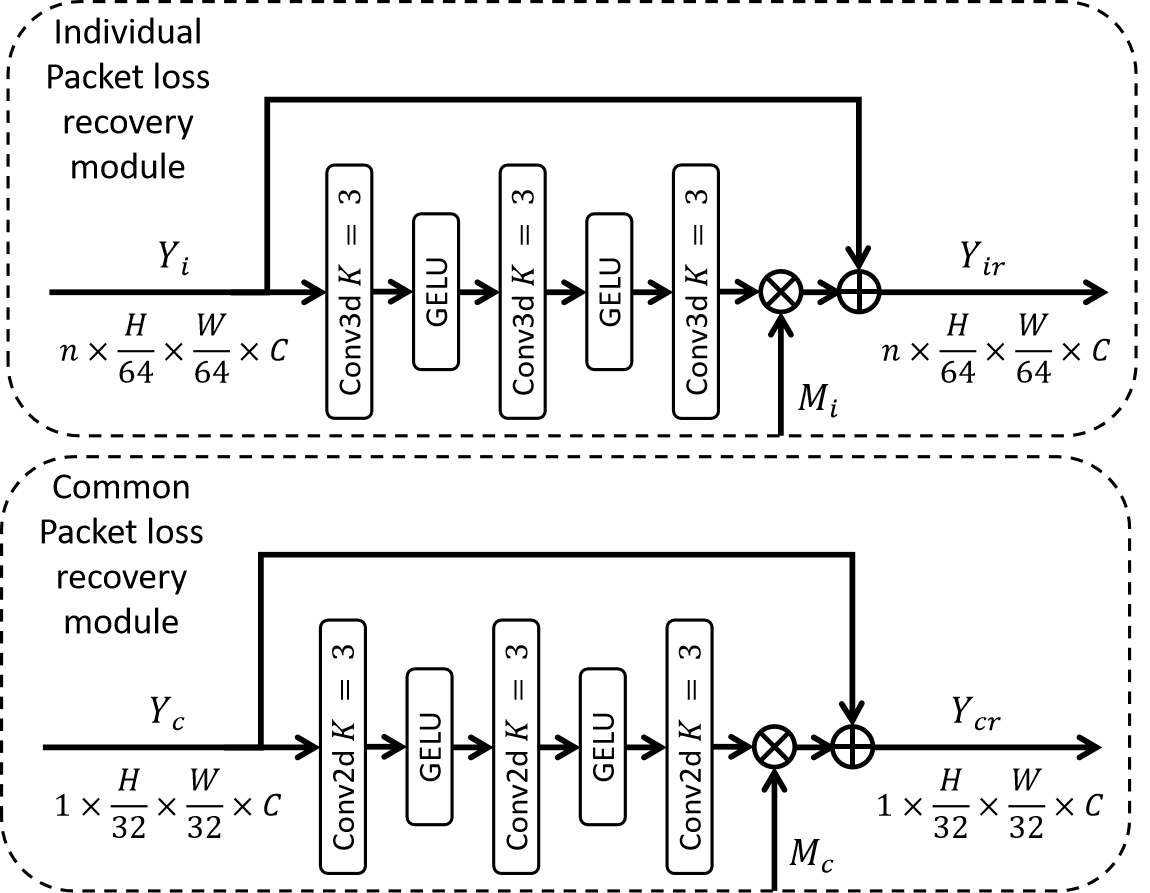}\\
	\caption{\footnotesize The architecture of the Packet loss recovery module.}
	\label{Packet loss recovery module}
\end{figure}

To address packet loss in communication, a Common Packet Loss Recovery Module and an Individual Packet Loss Recovery Module are developed, as illustrated in Fig. \ref{Packet loss recovery module}. In the Common Packet Loss Recovery Module, the semantic information vector \(Y_c\) with missing elements is processed through a convolutional network featuring a 2D convolutional layer with a \(3\times3\) kernel and an activation function. The output is multiplied by the packet-loss mask \(M_c\), producing a vector containing only the recovered missing semantic information, which is then added to the received \(Y_c\) to yield the complete semantic information vector \(Y_{cr}\). The Individual Packet Loss Recovery Module is designed similarly, differing only in its use of 3D convolution for three-dimensional input data, and thus is not described further. Unlike existing studies that often use channel state information as input at both encoding and decoding ends to design adaptive modules for harsh channels, real-world conditions make it difficult for the transmitter to obtain real-time channel state feedback from the receiver, leading to discrepancies between theoretical simulations and practical applications. Therefore, adaptive recovery is performed solely at the receiver based on received information, aligning more closely with practical scenarios.

\section{Experimental Results}
\subsection{Experimental Setup}
\subsubsection{Datasets}
The proposed MSTVSC is trained on the Vimeo-90k dataset \cite{ref30}, which contains 89,800 video clips, each comprising 7-frame sequences. The GOP size is set to 4, and the frames of the entire GOP are randomly cropped to 256 × 256 pixels. Subsequently, each frame’s data is normalized to the range [0, 1]. After training, the performance is evaluated using the HEVC test dataset \cite{ref31} and the UVG dataset \cite{ref32}. The following subsets are selected: Class A (2560 × 1440), Class B (1920 × 1080), and UVG (3840 × 2160). As analyzed in Section II, the UDP protocol is employed to handle semantic transmission tasks in this paper. Therefore, in simulations evaluating the impact of communication parameters on packet loss rate and semantic performance, the header length is set as follows: application layer (semantic segmentation) + transport layer (UDP) + network layer (IPv4) + data link layer (Ethernet header and FCS) = 3 + 8 + 20 + 18 = 49 bytes. Additionally, it is assumed that the semantic data to be transmitted is 155,520 bytes.
\subsubsection{Comparison Schemes}
The proposed MSTVSC scheme is compared with the CNN-based MDVSC scheme \cite{ref10} and classical separate source and channel coding schemes, including source coding with H.264 \cite{ref33} and H.265 \cite{ref34}, and channel coding with LDPC \cite{ref35}. In this paper, a 1/2 LDPC code implies that half of the total code length is utilized. Additionally, the impact of the presence or absence of the packet loss recovery module on combating packet loss, as well as the effect of varying the number of output channels \(C\) of the encoder on semantic performance, are also evaluated.
\subsubsection{Evaluation Metrics}
The end-to-end video transmission performance of the proposed MSTVSC model and other comparative schemes is validated using the widely adopted pixel-wise metric PSNR and the perceptual metric MS-SSIM \cite{ref36}. These two metrics can be calculated as follows:
\begin{equation}
	\begin{split}
\text{PSNR}(X, Y) = 10 \cdot \log_{10} \left( \frac{1}{\text{MSE}} \right)
	\end{split}
\end{equation}
\begin{equation}
	\begin{split}
		\operatorname{MSE}(\mathbf{X}, \mathbf{Y})=\frac{1}{w h} \sum_{i=0}^{w-1} \sum_{j=0}^{h-1}\|\mathbf{X}(i, j)-\mathbf{Y}(i, j)\|_{2},
	\end{split}
\end{equation}
where $w$ and $h$ represent the width and height of the input variable, respectively.
\begin{equation}\small
	\begin{split}
\operatorname{MS-SSIM}(X, Y)=\left[\mathcal{L}_{M}(X, Y)\right]^{\alpha_{M}} \cdot \prod_{j=1}^{M}\left[\mathcal{C}_{j}(X, Y) \cdot \mathcal{S}_{j}(X, Y)\right]^{\alpha_j},
	\end{split}
\end{equation}
where, \(M\) represents different dimensions, including luminance \(\mathcal{L}_{M}(\cdot)\), contrast \(\mathcal{C}_{j}(\cdot)\), and structure \(\mathcal{S}_{j}(\cdot)\), while \(\alpha_{M}\) and \(\alpha_j\) are the weights for the different terms.

Furthermore, the channel bandwidth ratio (CBR) is often used to measure bandwidth utilization efficiency and can be calculated as follows:
\begin{equation}
	\begin{split}
\text{CBR} = \frac{k}{m},
	\end{split}
\end{equation}
where \(m\) represents the dimension of the source data \(X_g\), and \(k\) is the sum of the dimensions of the vectors \(X_c\) and \(X_i\) to which \(X_g\) is mapped by the semantic encoder. The value \(k\) is also referred to as the channel bandwidth cost. It should be noted that, as the research focuses on digital communication systems, semantic data transmission requires quantization into bits. In this paper, a 3-bit quantization is applied to the encoding results of the semantic model.

\subsubsection{Model Training Details}
To balance video reconstruction quality in the absence of packet loss and high fault-tolerant decoding capability under packet loss scenarios, the training process is divided into two stages with different loss functions.

Training Stage I:
At this stage, the optimization objective is to train the model’s end-to-end video reconstruction capability without packet loss. Accordingly, the loss function for this stage is designed as:
\begin{equation}
	\begin{split}
\text{LOSS}_1 = \alpha \text{MSE}(X_g, Y_g) + \beta L_{aux_{MoE}},
	\end{split}
\end{equation}
where \(\text{MSE}\) is the mean squared error function, and \(L_{aux_{MoE}}\) is the auxiliary loss proposed in \cite{ref29} for training the MoE, ensuring that the gating function does not consistently select the same expert. Notably, since quantization lacks gradients, during training, quantization is implemented using the soft quantization method proposed in \cite{ref37}, defined as:
\begin{equation}
	\begin{split}
		Q(X) = X + q_n \quad \text{with} \quad q_n \sim U\left(-\frac{1}{2}, \frac{1}{2}\right),
	\end{split}
\end{equation}
where \(U(a, b)\) denotes a uniform distribution over the interval \([a, b]\). During validation and testing, the quantization operation is set to rounding.

Training Stage II:
After convergence in Stage I, the model’s ability to mitigate packet loss is trained. To align with practical scenarios, the packet loss channel is modeled as a semantic erasure channel. After passing through the packet loss channel, each element in the semantic information matrix may be dropped with a given packet loss rate \(P_L\), and dropped elements are filled with 0.

At this stage, the loss function is defined as:
\begin{equation}
	\begin{split}
		\text{LOSS}_2 = \text{LOSS}_1 + \gamma(\text{MSE}(X_i, Y_{ir}) + \text{MSE}(X_c, Y_{cr})),
	\end{split}
\end{equation}
where \(\text{MSE}(X_i, Y_{ir})\) and \(\text{MSE}(X_c, Y_{cr})\) are introduced to constrain the Individual Packet Loss Recovery Module and Common Packet Loss Recovery Module to reconstruct the original information sent by the transmitter as accurately as possible.

During training, the input \(P_L\) is an array, such as \([0.0, 0.3, 0.6, 0.9]\). Within the same batch, different GOPs randomly select one of these values as the parameter for the packet loss channel. Additionally, to preserve video quality in low packet loss scenarios, the encoder is frozen during this stage of training. This is because, when randomly selecting the packet loss channel parameter, a higher packet loss rate dominates the direction of loss reduction, which could lead the model to optimize toward combating high packet loss at the expense of reducing video detail and increasing redundant information. Instead, the goal is to train the decoder to capture the correlation of semantic information for frame recovery while maintaining video quality at low packet loss rates, rather than sacrificing detail for error redundancy. 

The Adam optimizer is utilized with a learning rate of \(1 \times 10^{-4}\) and a batch size set to 32. All implementations are completed using PyTorch on a single RTX A6000 GPU. To accommodate the computational requirements of different transmitting ends, models of varying sizes are designed, with their specific parameters listed in Tab. \ref{tab:model_size}. Unless otherwise specified in the Results Analysis subsection, the MSTVSC is assumed to use the mid-size model with a CBR of 0.0078. 

\begin{table}[h]
	\centering
	\begin{tabular}{>{\centering\arraybackslash}m{1.5cm} >{\centering\arraybackslash}m{1.5cm} >{\centering\arraybackslash}m{1.5cm} >{\centering\arraybackslash}m{1.5cm}}
		\toprule
		\textbf{model size} & \textbf{small} & \textbf{mid} & \textbf{large} \\ 
		\midrule
		$[N_1, N_2, N_3]$ & $[1, 1, 1]$ & $[2, 2, 2]$ & $[2, 2, 2]$ \\ 
		$[C_1, C_2, C_3]$ & $[24, 48, 72]$ & $[24, 48, 72]$ & $[48, 96, 144]$ \\ 
		$[N_4, N_5]$ & $[1, 1]$ & $[2, 2]$ & $[2, 2]$ \\ 
		$[C_4, C_5]$ & $[96, 120]$ & $[96, 120]$ & $[192, 240]$ \\ 
		$[N_6]$ & $[1]$ & $[2]$ & $[2]$ \\ 
		$[C_6]$ & $[144]$ & $[144]$ & $[288]$ \\ 
		$[C]$ & $[96]$ & $[96]$ & $[96]$ \\ 
		\bottomrule
	\end{tabular}
	\caption{Model Size Configurations}
	\label{tab:model_size}
\end{table}

\subsection{Results Analysis}

\subsubsection{The Variation of Packet Loss Rate and Semantic Performance with Channel Conditions and Communication Parameters}
\begin{figure}[!htbp]
	\centering
	\includegraphics[width=0.5\textwidth]{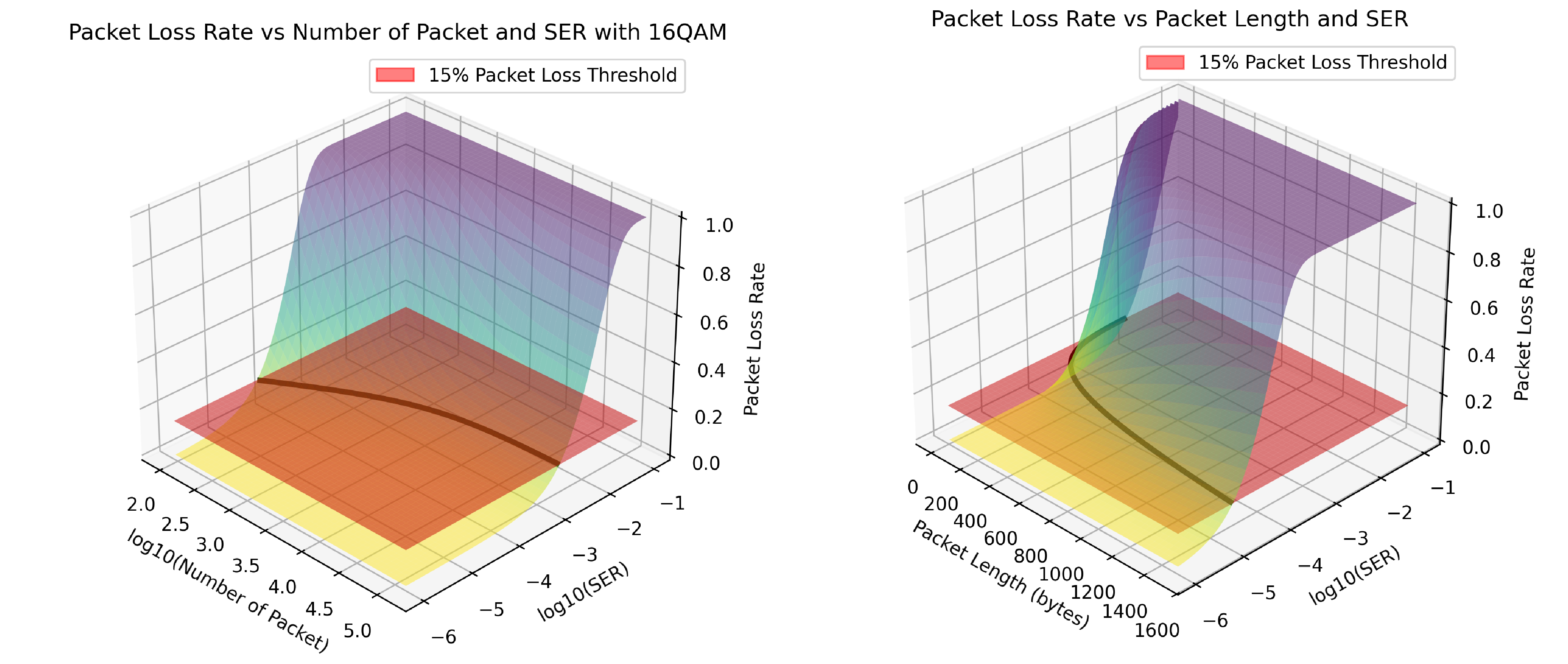}\\
	\caption{\footnotesize A three-dimensional surface plot illustrating the variation of packet loss rate with SER, number of packets and packet length is presented. A plane corresponding to a packet loss rate of 15\% is drawn in the figure, intersecting the surface along a black line. This black line represents the optimal choice of communication parameters for a given SER and packet loss rate threshold.}
	\label{3d1}
\end{figure}

The relationship between packet loss rate, channel conditions, and communication parameters, derived in Section II, is visualized in Fig. \ref{3d1}. For a fixed modulation order, the symbol error rate (SER) is determined by channel conditions. Given a specific SER, the packet loss rate is reduced as the number of packets increases and packet length decreases. Thus, the packet loss rate at the receiver can be adjusted by optimizing communication parameters. When a maximum packet loss rate threshold (e.g., 15\%) is set, the communication parameters yielding the least redundancy are identified by intersecting a threshold plane with the surface representing packet loss rate as a function of channel conditions and communication parameters. This intersection, shown as the black line in Fig. \ref{3d1} where the red threshold plane meets the surface, indicates the minimum \(L_{total}\) for various SERs when the packet loss rate meets or falls below the threshold. From these points, the optimal number of packets and corresponding packet length are determined. Consequently, the findings from Section II guide the optimization of communication parameters.


\begin{figure}[!htbp]
	\centering
	\includegraphics[width=0.48\textwidth]{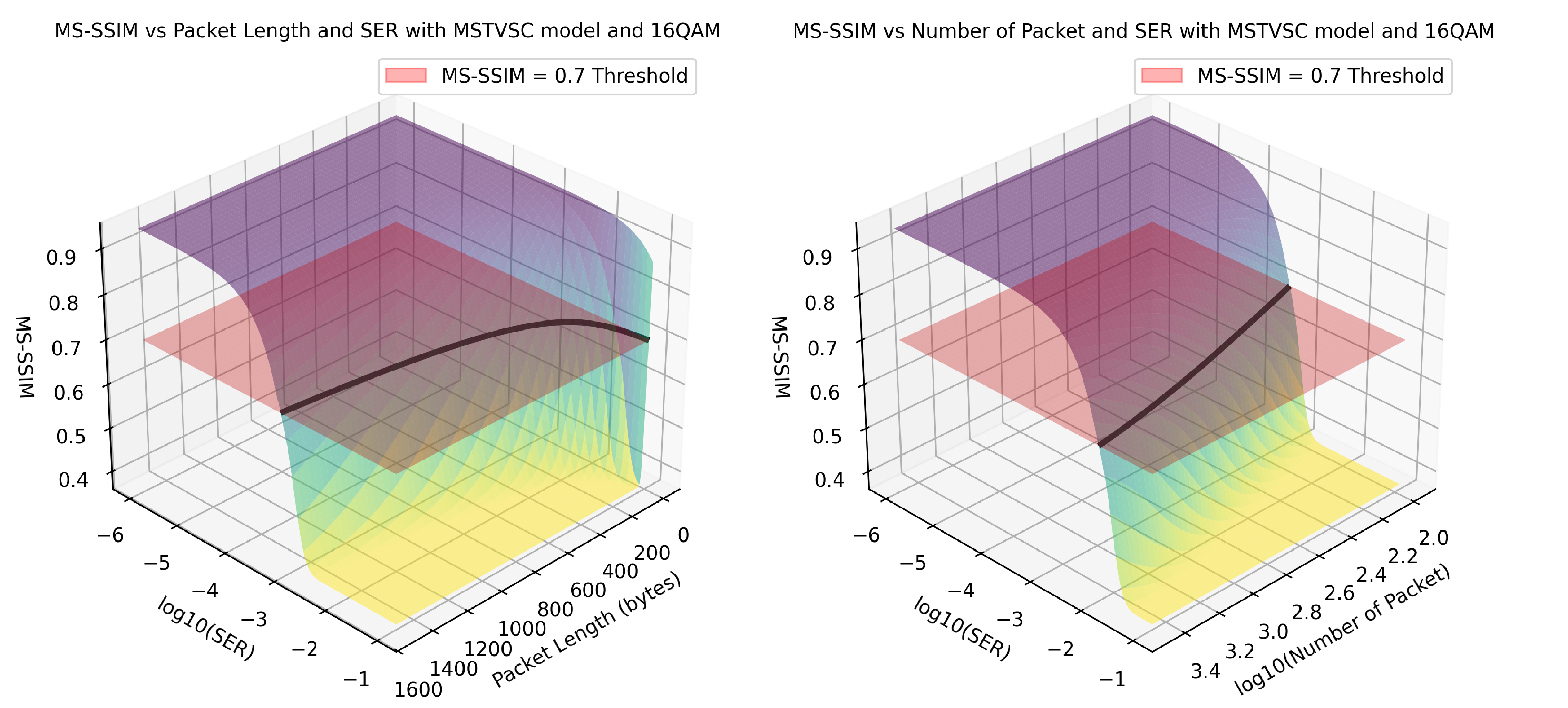}\\
	\caption{\footnotesize A three-dimensional surface plot illustrating the variation of semantic performance metrics MS-SSIM with SER, packet length and packet length is presented. Planes corresponding to MS-SSIM of 0.7 are drawn in the figure, intersecting the surface along black lines. These black lines represent the optimal choice of communication parameters for a given SER and packet loss rate threshold.}
	\label{3d2}
\end{figure}

The surface plot illustrating semantic performance as a function of channel conditions and communication parameters is presented in Fig. \ref{3d2}. This plot integrates data from Fig. \ref{packet_loss}, showing semantic performance variation with packet loss rate at 1920×1080 resolution, and Fig. \ref{3d1}, depicting packet loss rate variation with channel conditions and communication parameters. In practical semantic communication scenarios, packet length and redundancy ratio are adjusted based on the current SER and a predefined semantic performance threshold, such as an MS-SSIM of at least 0.7. This approach ensures the semantic performance threshold is met with minimal transmission volume under the UDP protocol.

\subsubsection{Performance at Different Packet Loss Rate}

\begin{figure*}[!htbp]
	\centering
	\includegraphics[width=0.9\textwidth]{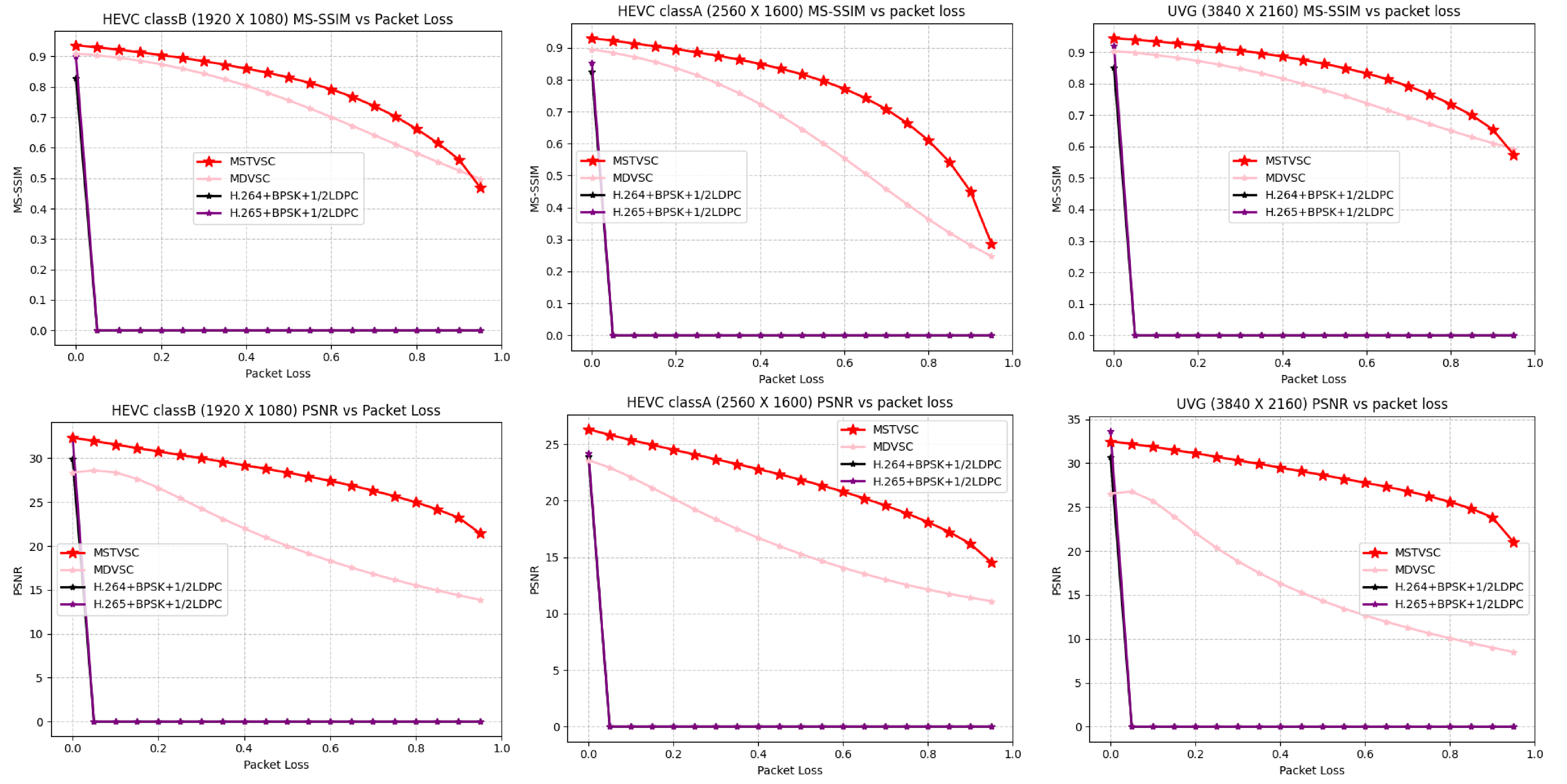}\\
	\caption{\footnotesize The relationship between MS-SSIM and PSNR performance and the packet loss rate under a packet loss channel is illustrated.}
	\label{packet_loss}
\end{figure*}

Performance variation curves for different schemes under packet loss at various resolutions are depicted in Fig. \ref{packet_loss}, with traditional methods using the channel robustness combination of BPSK + 1/2 LDPC. Due to the UDP protocol’s lack of retransmission, traditional schemes fail to decode and reconstruct video at the receiver when packet loss occurs. Across all resolutions and packet loss rates, MSTVSC significantly outperforms MDVSC in PSNR. For MS-SSIM, MSTVSC maintains superiority over MDVSC, with the advantage narrowing only as the packet loss rate nears 1. MDVSC’s performance declines rapidly with increasing packet loss, showing a steep, consistent linear slope. In contrast, MSTVSC’s MS-SSIM performance decreases more gradually at lower packet loss rates, with a smaller slope. Beyond a 60\% packet loss rate, the slope increases, indicating that MSTVSC effectively uses semantic information correlation to recover missing data at lower loss rates. However, at excessively high loss rates, insufficient un-lost data leads to significant semantic gaps and rapid performance decline. For PSNR, MSTVSC’s slope is also smaller than MDVSC’s, demonstrating greater tolerance to packet loss.

\subsubsection{Common and Individual Feature Extraction}
\begin{figure}[!htbp]
	\centering
	\includegraphics[width=0.49\textwidth]{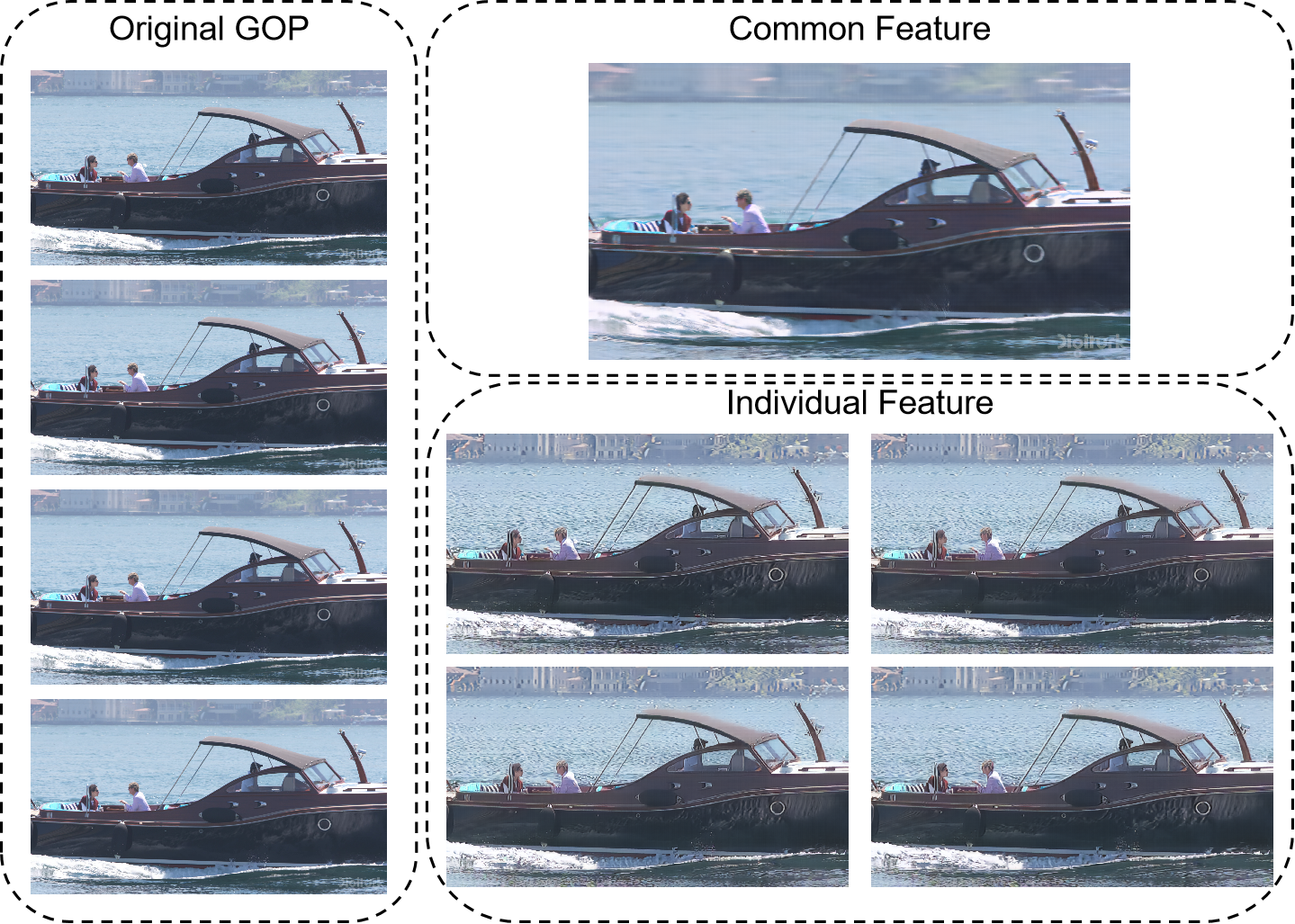}\\
	\caption{\footnotesize Visualization of Common and Individual Features.}
	\label{Feature}
\end{figure}

Reconstructed images of a GOP, decoded separately after common and individual feature extraction, are shown in Fig. \ref{Feature}. Common features primarily capture slowly varying information across the GOP, such as the boat hull and the logo in the bottom right corner, while rapidly changing elements like waves and background buildings appear blurred. In contrast, individual features effectively retain rapidly changing information, such as waves and background buildings, which are sharply defined, resembling a sharpening effect, while static elements like the logo are not preserved. This illustrates the effectiveness and distinct roles of common and individual feature decomposition, offering a visual insight into their contributions.

\subsubsection{Ablation Experiments}
\begin{figure}[!htbp]
	\centering
	\includegraphics[width=0.5\textwidth]{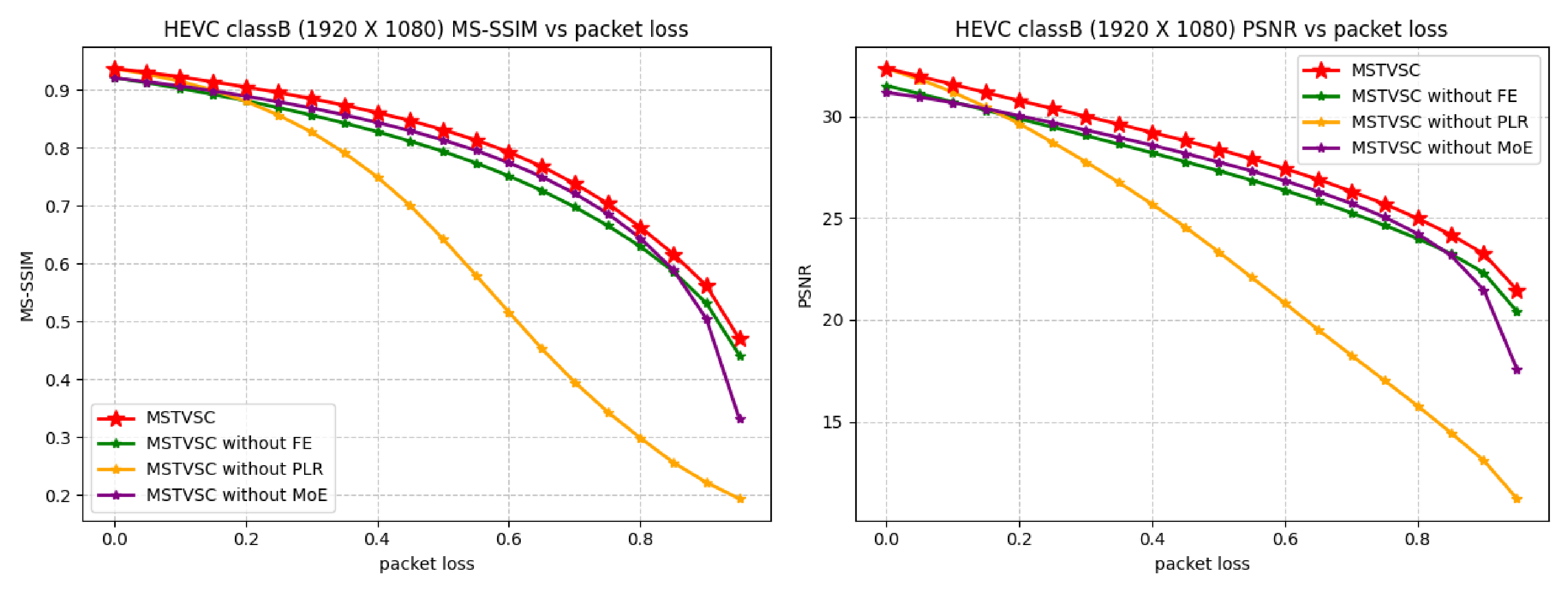}\\
	\caption{\footnotesize The ablation experiment result.}
	\label{xiaorong}
\end{figure}

Ablation experiment results at 1920×1080 resolution are depicted in Fig. \ref{xiaorong}. Compared to MSTVSC without the feature extraction (FE) system, the latter consistently underperforms, indicating that the common-individual separation module significantly improves semantic compression efficiency, enabling more effective semantic information compression at the same CBR.

When compared to MSTVSC without the packet loss recovery (PLR) system, the performance of the latter declines more rapidly as packet loss rate increases, ultimately falling below all other ablation counterparts. This highlights the PLR module’s effectiveness in mitigating semantic loss by recovering missing information using correlations between un-lost and lost semantics.

Additionally, MSTVSC without the Mixture of Experts (MoE) system is outperformed by the full MSTVSC, confirming that the MoE enhances semantic information extraction and reconstruction capabilities.

\section{Conclusion}

A semantic video communication system, termed MSTVSC, is proposed, designed to resist packet loss and ensure compatibility with UDP-based protocols. Correlated semantic information is dispersed across data segments through interleaving and segmentation at the application layer, thereby mitigating packet loss. At the receiver, semantic information is reassembled and de-interleaved using application-layer headers and pre-agreed indices. An expression for the packet loss rate, derived as a function of channel conditions and communication parameters, is established, and optimal parameters are determined based on the semantic performance variation curve obtained from MSTVSC testing. Missing information is recovered by a 3D CNN-based packet loss recovery module, which utilizes un-lost semantic data and a packet-loss mask matrix. To achieve high-quality video under high compression, redundancy is reduced through a common-individual feature separation method followed by downsampling of individual features. Semantic extraction and reconstruction are enhanced by a spatial-temporal codec based on the MoE Swin Transformer. For practical deployment, a lightweight codec employing spatial-temporal compression separation is developed. Compared with traditional methods H.264 and H.265, as well as the semantic method MDVSC, superior performance in video encoding-decoding, reconstruction, and robustness to packet loss is demonstrated by MSTVSC.


\begin{thebibliography}{37}
\bibliographystyle{IEEEtran}


	
\bibitem{ref1}
P. Zhang \textit{et al.}, ``Intellicise wireless networks from semantic communications: A survey, research issues, and challenges,'' \textit{IEEE Commun. Surveys Tuts.}, Early Access.

\bibitem{ref2}
W. Weaver, ``Recent contributions to the mathematical theory of communication,'' \textit{ETC: A Review of General Semantics}, vol. 10, no. 4, pp. 261--281, 1953.

\bibitem{ref3}
C. E. Shannon, ``A mathematical theory of communication,'' \textit{Bell Syst. Tech. J.}, vol. 27, no. 3, pp. 379--423, Jul. 1948.

\bibitem{ref4}
P. Zhang \textit{et al.}, ``Toward wisdom-evolutionary and primitive-concise 6G: A new paradigm of semantic communication networks,'' \textit{Engineering}, vol. 8, pp. 60--73, Jan. 2022.

\bibitem{ref5}
X. Luo, H.-H. Chen, and Q. Guo, ``Semantic communications: Overview, open issues, and future research directions,'' \textit{IEEE Wireless Commun.}, vol. 29, no. 1, pp. 210--219, Feb. 2022.

\bibitem{ref6}
K. Niu \textit{et al.}, ``A paradigm shift toward semantic communications,'' \textit{IEEE Commun. Mag.}, vol. 60, no. 11, pp. 113--119, Nov. 2022.



\bibitem{ref8}
C. Dong \textit{et al.}, ``Semantic communication system based on semantic slice models propagation,'' \textit{IEEE J. Sel. Areas Commun.}, vol. 41, no. 1, pp. 202--213, Jan. 2023.

\bibitem{ref9}
P. Zhang \textit{et al.}, ``Model division multiple access for semantic communications,'' \textit{Front. Inf. Technol. Electron. Eng.}, vol. 24, no. 6, pp. 801--812, Jun. 2023.

\bibitem{ref10}
Z. Bao \textit{et al.}, ``MDVSC---Efficient wireless model division video semantic communication,'' \textit{IEEE Internet Things J.}, vol. 12, no. 2, pp. 1109--1124, Jan. 2025.
\bibitem{ref7}
H. Xie, Z. Qin, G. Y. Li, and B.-H. Juang, ``Deep learning enabled semantic communication systems,'' \textit{IEEE Trans. Signal Process.}, vol. 69, pp. 2663--2675, 2021.
\bibitem{ref11}
N. Farsad, M. Rao, and A. Goldsmith, ``Deep learning for joint source-channel coding of text,'' in \textit{Proc. IEEE Int. Conf. Acoust., Speech, Signal Process. (ICASSP)}, Calgary, AB, Canada, Apr. 2018, pp. 2326--2330.

\bibitem{ref12}
X. Chen, J. Wang, L. Xu, J. Huang, and Z. Fei, ``A perceptually motivated approach for low-complexity speech semantic communication,'' \textit{IEEE Internet Things J.}, vol. 11, no. 12, pp. 22054--22065, Jun. 2024.

\bibitem{ref13}
Z. Weng and Z. Qin, ``Semantic communication systems for speech transmission,'' \textit{IEEE J. Sel. Areas Commun.}, vol. 39, no. 8, pp. 2434--2444, Aug. 2021.

\bibitem{ref14}
E. Bourtsoulatze, D. B. Kurka, and D. Gündüz, ``Deep joint source-channel coding for wireless image transmission,'' \textit{IEEE Trans. Cogn. Commun. Netw.}, vol. 5, no. 3, pp. 567--579, Sep. 2019.

\bibitem{ref15}
D. B. Kurka and D. Gündüz, ``DeepJSCC-f: Deep joint source-channel coding of images with feedback,'' \textit{IEEE J. Sel. Areas Inf. Theory}, vol. 1, no. 1, pp. 178--193, May 2020.

\bibitem{ref16}
S. Fan \textit{et al.}, ``A specific task-oriented semantic image communication system for substation patrol inspection,'' \textit{IEEE Trans. Power Del.}, vol. 39, no. 2, pp. 835--844, Apr. 2024.

\bibitem{ref17}
J. Dai \textit{et al.}, ``Nonlinear transform source-channel coding for semantic communications,'' \textit{IEEE J. Sel. Areas Commun.}, vol. 40, no. 8, pp. 2300--2316, Aug. 2022.



\bibitem{Qi2024}
X. Qi, N. Ma, Z. Bao, Y. Liu, C. Dong, and X. Xu, ``WAFI-VSC: Wireless adaptive frame interpolation video semantic communication,'' in \emph{Proc. 16th Int. Conf. Wireless Commun. Signal Process. (WCSP)}, Hefei, China, 2024, pp. 1503--1508.

\bibitem{Yang2025}
W. Yang, Z. Xiong, Y. Yuan, W. Jiang, T. Q. S. Quek, and M. Debbah, ``Agent-driven generative semantic communication with cross-modality and prediction,'' \emph{IEEE Trans. Wireless Commun.}, vol. 24, no. 3, pp. 2233--2248, Mar. 2025.

\bibitem{Jiang2023}
P. Jiang, C.-K. Wen, S. Jin, and G. Y. Li, ``Wireless semantic communications for video conferencing,'' \emph{IEEE J. Sel. Areas Commun.}, vol. 41, no. 1, pp. 230--244, Jan. 2023.

\bibitem{Samarathunga2024}
P. Samarathunga, Y. Ganearachchi, T. Fernando, A. Jayasingam, I. Alahapperuma, and A. Fernando, ``A semantic communication and vvc based hybrid video coding system,'' \emph{IEEE Access}, vol. 12, pp. 79202--79224, 2024.

\bibitem{Tong2025}
H. Tong, H. Li, H. Du, Z. Yang, C. Yin, and D. Niyato, ``Multimodal semantic communication for generative audio-driven video conferencing,'' \emph{IEEE Wireless Commun. Lett.}, vol. 14, no. 1, pp. 93--97, Jan. 2025.

\bibitem{Dai2023}
J. Dai \emph{et al.}, ``Toward adaptive semantic communications: Efficient data transmission via online learned nonlinear transform source-channel coding,'' \emph{IEEE J. Sel. Areas Commun.}, vol. 41, no. 8, pp. 2609--2627, Aug. 2023.

\bibitem{ref19}
M. Shi \textit{et al.}, ``DSCS: A decoupled semantic communication system for video conferencing,'' in \textit{Proc. 9th Int. Conf. Comput. Commun. Syst. (ICCCS)}, Xi'an, China, Apr. 2024, pp. 321--326.

\bibitem{ref20}
S. Wang \textit{et al.}, ``Wireless deep video semantic transmission,'' \textit{IEEE J. Sel. Areas Commun.}, vol. 41, no. 1, pp. 214--229, Jan. 2023.

\bibitem{ref21}
T.-Y. Tung and D. Gündüz, ``DeepWiVe: Deep-learning-aided wireless video transmission,'' \textit{IEEE J. Sel. Areas Commun.}, vol. 40, no. 9, pp. 2570--2583, Sep. 2022.
\bibitem{Tian2025}
Y. Tian, J. Ying, Z. Qin, Y. Jin and X. Tao, "Synchronous Multi-Modal Semantic Communication System With Packet-Level Coding," \emph{IEEE Trans. Wireless Commun.}, vol. 24, no. 5, pp. 3684-3697, May 2025.






















\bibitem{ref22}
H. Zhang \textit{et al.}, ``Diffusion-based wireless semantic communication for VR image,'' in \textit{Proc. IEEE/CIC Int. Conf. Commun. China (ICCC Workshops)}, Hangzhou, China, Aug. 2024, pp. 639--644.

\bibitem{ref23}
X. Liu \textit{et al.}, ``A semantic communication system for point cloud,'' \textit{IEEE Trans. Veh. Technol.}, vol. 74, no. 1, pp. 894--910, Jan. 2025.



\bibitem{ref26}
L. Teng, W. An, C. Dong, and X. Xu, ``sDMCM---Semantic digital modulation constellation mapping scheme for semantic communication,'' \textit{IEEE Internet Things J.}, Early Access.


\bibitem{ref27}
Z. Liu \textit{et al.}, ``Swin transformer: Hierarchical vision transformer using shifted windows,'' in \textit{Proc. IEEE/CVF Int. Conf. Comput. Vis. (ICCV)}, Montreal, QC, Canada, Oct. 2021, pp. 9992--10002.

\bibitem{ref25}
K. Yang, S. Wang, J. Dai, X. Qin, K. Niu, and P. Zhang, ``SwinJSCC: Taming Swin Transformer for Deep Joint Source-Channel Coding,'' \textit{IEEE Trans. Cogn. Commun. Netw.}, vol. 11, no. 1, pp. 90--104, Feb. 2025.

\bibitem{ref28}
Z. Liu \textit{et al.}, ``Video Swin transformer,'' in \textit{Proc. IEEE/CVF Conf. Comput. Vis. Pattern Recognit. (CVPR)}, New Orleans, LA, USA, Jun. 2022, pp. 3192--3201.

\bibitem{ref29}
D. Lepikhin \textit{et al.}, ``GShard: Scaling giant models with conditional computation and automatic sharding,'' \textit{arXiv preprint}, 2020. [Online]. Available: https://arxiv.org/abs/2006.16668

\bibitem{ref30}
T. Xue, B. Chen, J. Wu, D. Wei, and W. T. Freeman, ``Video enhancement with task-oriented flow,'' \textit{Int. J. Comput. Vis.}, vol. 127, no. 8, pp. 1106--1125, Aug. 2019.

\bibitem{ref31}
F. Bossen \textit{et al.}, ``Common test conditions and software reference configurations,'' Int. Organ. Standard., Geneva, Switzerland, document JCTVC-L1100, 2013.

\bibitem{ref32}
A. Mercat, M. Viitanen, and J. Vanne, ``UVG dataset: 50/120fps 4K sequences for video codec analysis and development,'' in \textit{Proc. 11th ACM Multimedia Syst. Conf.}, Istanbul, Turkey, Jun. 2020, pp. 297--302.

\bibitem{ref33}
T. Wiegand, G. J. Sullivan, G. Bjøntegaard, and A. Luthra, ``Overview of the H.264/AVC video coding standard,'' \textit{IEEE Trans. Circuits Syst. Video Technol.}, vol. 13, no. 7, pp. 560--576, Jul. 2003.

\bibitem{ref34}
G. J. Sullivan, J.-R. Ohm, W. Han, and T. Wiegand, ``Overview of the high efficiency video coding (HEVC) standard,'' \textit{IEEE Trans. Circuits Syst. Video Technol.}, vol. 22, no. 12, pp. 1649--1668, Dec. 2012.

\bibitem{ref35}
T. J. Richardson and S. Kudekar, ``Design of low-density parity check codes for 5G new radio,'' \textit{IEEE Commun. Mag.}, vol. 56, no. 3, pp. 28--34, Mar. 2018.

\bibitem{ref36}
Z. Wang, E. P. Simoncelli, and A. C. Bovik, ``Multiscale structural similarity for image quality assessment,'' in \textit{Proc. 37th Asilomar Conf. Signals, Syst., Comput.}, Pacific Grove, CA, USA, Nov. 2003, pp. 1398--1402.

\bibitem{ref37}
J. Ballé, V. Laparra, and E. P. Simoncelli, ``End-to-end optimized image compression,'' \textit{arXiv preprint}, 2016. [Online]. Available: https://arxiv.org/abs/1611.01704.
	


\end{thebibliography}
\end{document}